\documentclass[preprint,twocolumn,amsmath,amssymb]{jpsj3} 

\usepackage{bm} 

\usepackage{enumerate}
\usepackage{graphicx,color}
\usepackage{amsmath,amsfonts,amsthm,amssymb,slashbox} 
\usepackage{dcolumn}
\usepackage{multicol}
\usepackage{ulem}



\def\runauthor{Kurita, Yamaji and Imada}

\title{
Topological Insulators from Spontaneous Symmetry Breaking Induced by Electron Correlation on Pyrochlore Lattices
} 
 
\author{  
Moyuru \textsc{Kurita}$^{1}$, Youhei \textsc{Yamaji}$^{1,2}$\thanks{present address: Department of Physics and Astronomy, Rutgers University, 136 Frelinghuysen Road, Piscataway, NJ 08854-8019, USA}, and Masatoshi \textsc{Imada}$^{1,2}$}
 
\inst{
$^{1}$ Department of Applied Physics,  
 University of Tokyo, 7-3-1 Hongo, Bunkyo-ku, Tokyo 113-8656 \\  
$^{2}$ Japan Science and Technology Agency, CREST, Honcho, Kawaguchi, Saitama 332-0012, Japan 
} 
 
\recdate{February 25, 2011} 
 
\abst{We study an extended Hubbard model with the nearest-neighbor Coulomb interaction on the pyrochlore lattice at half filling.
An interaction-driven insulating phase with nontrivial $Z_2$ invariants emerges at the Hartree-Fock mean-field level in the phase diagram.
This topological insulator phase competes with other ordered states and survives in a parameter region surrounded by a semimetal, antiferromagnetic and charge ordered insulators.
The symmetries of these phases are group-theoretically analyzed.
We also show that the ferromagnetic interaction enhances the stability of the topological phase.
 } 

\kword{%
topological insulator, strongly correlated electron system, spontaneous symmetry breaking, pyrochlore lattice, extended Hubbard model} 
 
\begin{document} 
\maketitle 

\section{Introduction}
The discovery of the quantum Hall effect \cite{cit:PRL45_494} and theoretical approaches to the effect \cite{cit:PRB23_5632,cit:PRL49_405} have revealed the importance of topological properties of electron bands.
In the case of the quantum Hall states, the topological properties of systems are characterized by an integer called Chern number, which is physically measured as the quantized Hall conductance.
A non-zero Chern number requires a breaking of the time-reversal symmetry typically realized by applying strong magnetic fields on a two-dimensional system.
The quantized Hall conductance is also topologically protected by the structures of the edge states \cite{cit:PRB48_11851}.

Recently, topological classification of the time-reversal-invariant band insulators in two and three spatial dimensions has been made in seminal works \cite{cit:PRL95_146802,cit:PRB74_195312,cit:PRL98_106803,cit:PRB75_121306,cit:PRB79_195322}.
The classification can be completed by using the knowledge of the single-particle wave functions alone.
Two-dimensional band insulators are characterized by a single $Z_2$ invariant, which is equal to the number of the gapless edge states modulo 2 protected by the time-reversal symmetry.
Insulators with nontrivial $Z_2$ invariant namely $Z_2$ invariant $= 1$, are termed as topological insulators (TIs) or quantum spin Hall (QSH) insulators.
In three dimensions, four $Z_2$ invariants $(\nu_0; \nu_1\nu_2\nu_3)$ with $\nu_\alpha = 0$ or $1$ characterize either strong topological insulators (STIs), weak topological insulators (WTIs), or trivial band insulators.
STIs have $\nu_0 = 1$ while WTIs have some of the $(\nu_1\nu_2\nu_3)$ differ from $0$ and $\nu_0 = 0$.
Trivial insulators are characterized by the index $(0;000)$.

Both in two and three dimensions, the topological phases are typically realized in the systems with strong spin-orbit (SO) interaction.\cite{cit:Science314_1757,cit:Science318_766,cit:Nature452_970}
On the other hand, it has been suggested that the extended Hubbard model on the honeycomb lattice can generate an effective SO interaction from a spontaneous symmetry breaking at the Hartree-Fock mean-field level and results in the QSH phase \cite{cit:PRL100_156401}.
Similar phenomenon has also been proposed on the kagom\'{e} and diamond lattices \cite{cit:PRB82_075125,cit:PRB79_245331}.
A common property is that the lattice models which are semimetals in the single particle problem may have topologically nontrivial insulator phases caused by the Coulomb interaction.
Therefore, these states are called as topological Mott insulators (TMIs) and we will adopt this terminology in this paper.
We note that TMI has also been used in a different context \cite{cit:PRB78_125316,cit:NPhys6_376}.

In this paper, we study the TMI on the pyrochlore lattice.
Systems with pyrochlore lattices have been extensively studied because they provide us with various quantum effects.
Recently, it has been shown that a tight-binding model with a SO interaction and a lattice distortion generates the STI on the pyrochlore lattice \cite{cit:PRL103_206805}.
Specifically, the SO interaction yields the STI with $(1;000)$ at half filling.
Possibility of STI in an actual pyrochlore iridate has also been suggested \cite{cit:PRB82_085111}.
In our paper, we propose that a tight-binding model on the pyrochlore lattice becomes STI at half filling when the Coulomb interaction is switched on at the Hartree-Fock mean-field level.
In fact a STI state emerges realistically solely from an appropriate nearest neighbor interaction in contrast to other proposed TMI, where stronger next-nearest neighbor interactions are assumed.
We mainly focus on half filling, where the STI has a chance to emerge as we shall see.
In addition, competitions between other possible symmetry broken states are important at half filling as well.
The system is a semimetal at half and quarter filling without the Coulomb interaction.

This paper is organized as follows: In $\S$\ref{sec:Microscopic Hamiltonian and Topological Phases without Electron Correlations}, we introduce a microscopic tight-binding Hamiltonian on the pyrochlore lattice.
Its ground state becomes TI when the SO interaction is switched on.
We confirm it from the $Z_2$ invariants.
In $\S$\ref{sec:Interaction Effects by Hartree-Fock Mean-Field Calculation}, we consider electron correlation effects by employing an extended Hubbard model with a nearest-neighbor interaction and without the SO interaction.
The Hartree-Fock mean-field calculation gives a phase diagram which contains TI in a region of the parameter space even without the SO interaction.

\section{Microscopic Hamiltonian and Topological Phases without Electron Correlations}\label{sec:Microscopic Hamiltonian and Topological Phases without Electron Correlations}

\begin{figure}
	\centering
	\includegraphics{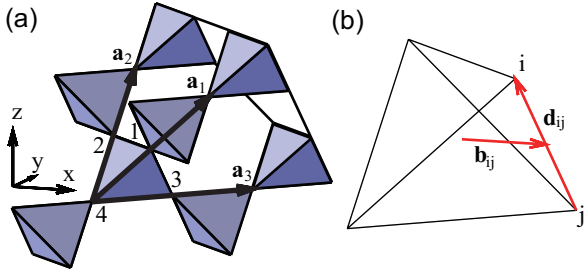}
	\caption{(Color online) (a) Pyrochlore lattice with its unit translation vectors forming a face-centered cubic(fcc) Bravais lattice. The unit structure is a shaded tetrahedron consiting of sites denoted by $1,2,3,4$. (b) Unit cell with graphical definitions of $\bm{b}_{ij}$ and $\bm{d}_{ij}$.}
	\label{fig:pyrochlore}
\end{figure}

\begin{figure}
	\centering
	\includegraphics{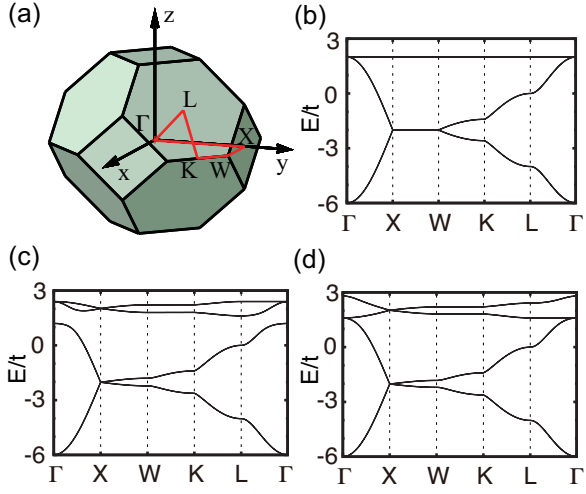}
	\caption{(Color online) (a) First Brillouin zone of fcc lattice with high-symmetry lines and points. (b) Band structure of the Hamiltonian $H_{0}$ in eq. (\ref{eq:NNhopping}) or (\ref{eq:H0k}). (c) and (d) Band structure of the Hamiltonian $H_{0} + H_{\mathrm{SO}}$ in eqs.(\ref{eq:NNhopping}) and (\ref{eq:HSO}) for $\lambda = 0.1t$ and $\lambda = -0.1t$, respectively}
	\label{fig:bulk_band}
\end{figure}

In this section, we introduce a microscopic Hamiltonian on the pyrochlore lattice that contains a topologically nontrivial state.
We consider in this section non-interacting systems with the SO coupling taken into account to provide a basis of studies on electron correlation effects in the next section.
We show that a STI state emerges solely from an appropriate nearest neighbor hopping due to SO coupling in 3D systems in contrast to the previous studies \cite{cit:PRL98_106803,cit:PRL103_206805}.
The pyrochlore structure is represented by an fcc Bravais lattice with a tetrahedral unit cell as shown in Fig. \ref{fig:pyrochlore}(a).
We choose the unit cell vectors to be
\begin{equation}
	\bm{a}_{1}=(2,0,2),\ \bm{a}_{2}=(0,2,2)\  \textrm{and} \ \bm{a}_{3}=(2,2,0).\label{eq:Bravais}
\end{equation}
Then the reciprocal lattice vectors are given by
\begin{eqnarray}
	\bm{b}_{1} &=&\frac{\pi}{2}(1,-1,1),\ \bm{b}_{2}=\frac{\pi}{2}(-1,1,1) \ \textrm{and} \ \notag \\
	 \bm{b}_{3} &=& \frac{\pi}{2}(1,1,-1).
\end{eqnarray}
The first Brillouin zone is shown in Fig. \ref{fig:bulk_band}(a) with high-symmetry lines and points.

Our starting point is the tight-binding Hamiltonian
\begin{equation}
	H_0 = -t\sum_{\langle ij \rangle \sigma}(c^{\dagger}_{i\sigma}c_{j\sigma}+h.c.), \label{eq:NNhopping}
\end{equation}
where $c^{\dagger}_{i\sigma}$ creates an electron with spin $\sigma$ on the site $i$ and $\langle ij\rangle$ denotes a nearest neighbor bond.
In the momentum space, eq. (\ref{eq:NNhopping}) becomes 
\begin{eqnarray}
	H_0 = \sum_{\bm{k}nm\alpha\beta}c_{\bm{k}n\alpha}^{\dagger}H_0(\bm{k})_{n\alpha m\beta}c_{\bm{k}m\beta}
\end{eqnarray}
with
\begin{eqnarray}
	H_0(\bm{k}) &=& -2tA(\bm{k})\otimes\sigma_0, \nonumber \\
	A(\bm{k}) &=&
	\left( \begin{array}{cccc}
	0 & \cos(k_x-k_y) & \cos(k_y-k_z) & \cos(k_x+k_z) \\
	 & 0 & \cos(k_x-k_z) & \cos(k_y+k_z) \\
	 &  & 0 & \cos(k_x+k_y) \\
	 &  &  & 0 \\
	\end{array} \right), \nonumber \\ \label{eq:H0k}
\end{eqnarray}
where $c_{\bm{k}n\sigma}^{\dagger}$ is the creation operator of a Bloch electron on the sublattice $n = (1,2,3,4)$ with the wave vector $\bm{k}$ and spin $\sigma$, and $\sigma_0$ is the identity matrix in spin space.
The lower triangle of the matrix $A(\bm{k})$ should be filled so that the matrix is Hermitian.
The spectrum of eq. (\ref{eq:H0k}), shown in Fig. \ref{fig:bulk_band}(b), is described analytically as
\begin{eqnarray}
	E_{(1,2)}(\bm{k}) & = & -2t\left(1\pm\sqrt{1+ B(\bm{k})}\right), \nonumber \\
	B(\bm{k}) &=& \cos2k_{x}\cos2k_{y}+\cos2k_{x}\cos2k_{z} \nonumber \\
			&+& \cos2k_{y}\cos2k_{z}
\end{eqnarray}
and
\begin{eqnarray}
	E_{(3,4)}(\bm{k}) & = & 2t.
\end{eqnarray}
Here, $E_{(2)}(\bm{k})$ touches the two flat bands $E_{(3,4)}(\bm{k})$ quadratically at the $\Gamma$ point at $E/t = 2.0$, while $E_{(1)}(\bm{k})$ and $E_{(2)}(\bm{k})$ are degenerate at the diagonal line of the square faces of the Brillouin zone, namely along the line WX at $E/t = -2.0$.

Next, we introduce a spin-dependent nearest-neighbor hopping which is crucial for the emergence of the topological states.
We consider the SO interaction of the form
\begin{equation}
	H_{\mathrm{SO}} = \sqrt{2}\lambda\sum_{\langle ij \rangle \alpha\beta}\left(i c_{i\alpha}^{\dagger}\frac{\bm{b}_{ij}\times\bm{d}_{ij}}{|\bm{b}_{ij}\times\bm{d}_{ij}|}\cdot\bm{\sigma}_{\alpha\beta}c_{j\beta} + h.c. \right), \label{eq:HSO}
\end{equation}
where $\lambda$ is the SO coupling strength, $\bm{\sigma}$ is the Pauli spin matrix, $\bm{b}_{ij}$ is the vector from the center of a tetrahedron to the midpoint of the bond $\langle ij \rangle$ constituting the tetrahedron edge and $\bm{d}_{ij}$ is the vector connecting the site $j$ to $i$.
Vectors $\bm{b}_{ij}$ and $\bm{d}_{ij}$ are graphically shown in Fig. \ref{fig:pyrochlore}(b).
Since each bond participates in forming only a single tetrahedron, $\bm{b}_{ij}$ is uniquely determined when a bond $\langle ij \rangle$ is specified.
We show that this form of the SO interaction is justified from a group theoretical argument in $\S$ \ref{subsec:Symmetry of pyrochlore lattice}. 
We note that this SO interaction may be viewed as a three dimentional generalization of the SO interaction introduced on the kagom\'{e} lattices in Refs.\citen{cit:PRB82_075125,cit:PRA82_053605}, where the starting point of $\bm{b}_{ij}$ is the center of the triangle.
However, there exists a clear distinction between the kagom\'{e} and pyrochlore lattices:
In the case of the kagom\'{e} lattice, $\bm{b}_{ij}\times\bm{d}_{ij}$ is perpendicular to the kagom\'{e} plane for all the bonds $\langle ij \rangle$.
On the other hand, in our three-dimentional system, the vectors ${\bm{b}_{ij}}\times\bm{d}_{ij}$ are not even coplaner, which leads to the absence of a unitary or an antiunitary transformation (ex. a spin rotation or a time-reversal transformation) that transforms $\lambda$ to $-\lambda$.
Therefore, the spectrum of $H_0 + H_{\mathrm{SO}}$ with positive $\lambda$ and negative $\lambda$ differs each other as shown in Figs. \ref{fig:bulk_band}(c) and (d).
A similar behavior was reported in the case of the next nearest neighbor SO interaction \cite{cit:PRL103_206805}.
In the present case, $H_{\mathrm{SO}}$ with positive $\lambda$ opens a gap at half filling while negative $\lambda$ does not.

Now we study the topological invariants of the insulating states derived from $H_0 + H_{\mathrm{SO}}$.
We can easily evaluate $(\nu_0;\nu_1\nu_2\nu_3)$, since our Hamiltonian has the inversion symmetry \cite{cit:PRB76_045302}.
The invariants can be calculated from the parity eigenvalues $\xi_{2m}(\bm{\Gamma}_i)$ of the $2m$-th occupied energy band at the $8$ time-reversal-invariant momenta (TRIM) $\bm{\Gamma}_i$ which satisfy $\bm{\Gamma}_i = \bm{\Gamma}_i + \bm{G}$, with an appropriate reciprocal lattice vector $\bm{G}$.
The $8$ TRIM in our system can be expressed as
\begin{equation}
 \bm{\Gamma}_{i=(n_{1}n_{2}n_{3})}=(n_{1}\bm{b}_{1}+n_{2}\bm{b}_{2}+n_{3}\bm{b}_{3})/2,
\end{equation}
with $n_{j} = 0,1$, where $\bm{b}_{i}$ are the primitive reciprocal lattice vectors.
Then $\nu_\alpha$ is obtained by the product
\begin{eqnarray}
	(-1)^{\nu_0} & = & \prod_{n_{j}=0,1}\delta_{n_{1}n_{2}n_{3}}, \nonumber \\
	(-1)^{\nu_{i=1,2,3}} & = & \prod_{n_{j \neq i}=0,1;n_{i}=1}\delta_{n_{1}n_{2}n_{3}},
\end{eqnarray}
where $\delta_{i}=\prod_{m=1}^{N}\xi_{2m}(\bm{\Gamma_{i}})$, with the parity eigenvalue $\xi_{2m}$ given below and $N$ being the number of Kramers pairs of occupied bands which are paired by the time reversal.
If we select the site 4 in Fig. \ref{fig:pyrochlore}(a) as the center of the inversion then the parity operator $P$ acts as $Pc_{\bm{k}n\sigma} = c_{-\bm{k}n\sigma}$ on the annihilation operator in the $\bm{k}$ space.
Since $c_{-\bm{\Gamma}_{i}n\sigma} = c_{\bm{\Gamma}_{i}n\sigma}e^{-i\bm{a}_{n}\cdot\bm{\Gamma_{i}}}$ is satisfied at the $8$ TRIM with $\bm{a}_{n}(n=1,2,3)$ being defined by eq. (\ref{eq:Bravais}) and $\bm{a}_{4} = \bm{0}$, parity eigenvalues for the eigenvectors of $H_0(\bm{\Gamma_i})+H_{\mathrm{SO}}(\bm{\Gamma_i})$ are nothing but the eigenvalues of the $8\times 8$ matrix $P_{\bm{\Gamma_i}} = \mathrm{diag}(e^{-i\bm{a}_{1}\cdot\bm{\Gamma_{i}}},e^{-i\bm{a}_{2}\cdot\bm{\Gamma_{i}}},e^{-i\bm{a}_{3}\cdot\bm{\Gamma_{i}}},e^{-i\bm{a}_{4}\cdot\bm{\Gamma_{i}}})\otimes\sigma_0$, where $\mathrm{diag}(a,b,c,d)$ denotes a diagonal matrix whose elements are $a,b,c$ and $d$.
Here, $H_{\mathrm{SO}}(\bm{k})$ is defined so as to satisfy $H_{\mathrm{SO}} = \sum_{nm\alpha\beta}c_{\bm{k}n\alpha}^{\dagger}H_{\mathrm{SO}}(\bm{k})_{n\alpha m\beta}c_{\bm{k}m\beta}$.
Now $\xi_{2m}(\bm{\Gamma}_i)$ is obtained by
\begin{eqnarray}
	P_{\bm{\Gamma}_i}\bm{u}_{\bm{\Gamma}_{i}2m} =  \xi_{2m}(\bm{\Gamma}_i) \bm{u}_{\bm{\Gamma}_{i}2m},
\end{eqnarray}
where $\bm{u}_{\bm{\Gamma}_{i}2m}$ is the $2m$-th eigenvector of the matrix $H_0(\bm{\Gamma_i})+H_{\mathrm{SO}}(\bm{\Gamma_i})$.

We first consider the case at half filling, where the gap opens for $\lambda > 0$.
We find that the index is $(1;000)$, which indicates that the phase is a STI for $\lambda > 0$.

On the other hand, at quarter filling, there exists a subtlety.
We observe that Dirac type band crossings are present even for nonzero $\lambda$ irrespective of its sign at three $\bm{X}$ points at $E/t=-2$.
At quarter filling, the Fermi level is located at this Dirac point at $E/t=-2$.
Therefore, even for nonzero $\lambda$, the gap does not open that makes $Z_2$ invariants ill defined in contrast to the case at half filling.
To make them well defined, the degeneracy at the Fermi level must be removed.
Therefore, at quarter filling, each filled band must be completely separated from the empty bands to study physics of topological insulators.
For that purpose, one may introduce a small lattice distortion labeled by $l = 1, 2, 3, 4$ as described by the Hamiltonian
\begin{equation}
	H_{\mathrm{dis}}^{l} = -d\sum_{\langle ij \rangle \sigma}\left( \mathrm{sgn}(l:ij)c^{\dagger}_{i\sigma}c_{j\sigma} + h.c. \right),
\end{equation}
where $d$ is the strength of the distortion, and $\mathrm{sgn}(l:ij)$ is $+$ when the sublattice label of the site $i$ or $j$ is $l$ and $-$, otherwise.
Non-zero $d$ opens a gap at all the Dirac points.
Such a distortion may be realized by trigonally uniaxial pressure, for instance in the direction (1,1,1).

In fact at quarter filling, $d = 0$ is the critical point separating STI and WTI.
The system is STI for positive $d$ and WTI for negative $d$.
This can be understood since $H_0 + H_{\mathrm{dis}}^l$ with $d=-t$ makes independent stackings of kagom\'{e} lattices.
We find that $(\nu_{1}\nu_{2}\nu_{3})$ depends solely on $l$ and not on the sign of $d$.
Therefore our Hamiltonian can yield 8 different topological phases depending on the sign of $d$ and $l=1,2,3,4$.
The vector $\bm{G}_{\nu}\equiv\sum_{i=1,2,3}\nu_{i}\bm{b}_i$, which can be defined independently of the choice of ${\bm{b}_i}$ \cite{cit:PRL98_106803}, is parallel to the axis passing through the center of the tetrahedron and the site $l$ in Fig. \ref{fig:pyrochlore}(a).
This result is consistent with the argument in Ref.\citen{cit:PRL98_106803}, which suggests that the WTI is interpreted as a pile of two-dimensional QSH states stacked in the $\bm{G}_{\nu}$ direction.
The topological invariants obtained here are identical to those in Ref.\citen{cit:PRL103_206805}, when the next nearest neighbor SO interaction is considered.
Although we found that our Hamiltonian and their Hamiltonian are linked without closing a gap, the significance of the present result is that the nearest-neighbor SO coupling in our Hamiltonian can be reproduced as an effective hopping matrix solely by the electron correlation of a realistic value, as we will discuss in the next section.

\begin{figure}
	\centering
	\includegraphics{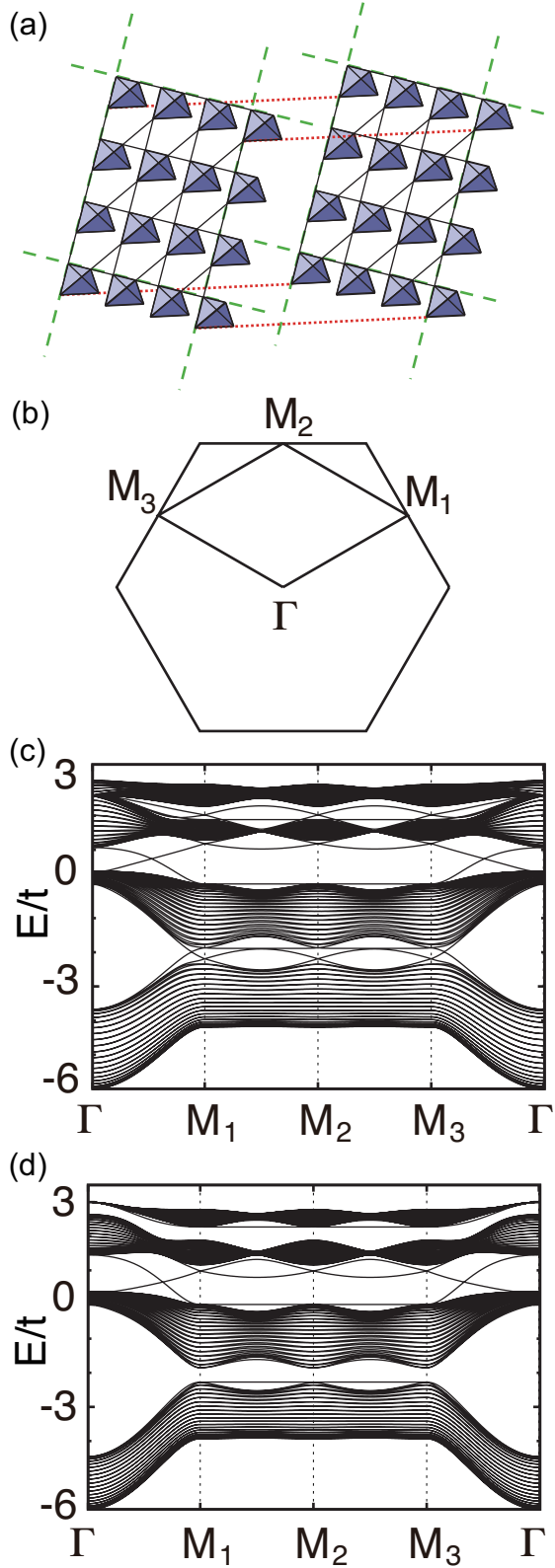}
	\caption{(Color online) (a) Slab of pyrochlore lattice. We take periodic  boundary conditions along the dashed (green) lines and open boundary condition along the dotted (red) lines. (b) Surface Brillouin zone specified with high-symmetry points and lines. (c) and (d) Band structures of the lattice Hamiltonian $H_0 + H_{\mathrm{SO}} + H_{\mathrm{dis}}^l$ on the slab with $\lambda = 0.2t$ and $d=\pm 0.1t$, respectively. The sublattice index $l$ is chosen in order that the axis passing through the center of the tetrahedron and the sublattice $l$ become perpendicular to the slab.}
	\label{fig:slab_all}
\end{figure}

The topological nontriviality of our system can also be seen by studying the surface states of the slab geometry, where we take the periodic boundary conditions along the two spatial directions and an open boundary condition along the other direction as shown in Fig. \ref{fig:slab_all}(a).
Figures \ref{fig:slab_all}(c) and (d) show the numerical diagonalization of the lattice Hamiltonian $H_0 + H_{\mathrm{SO}} + H_{\mathrm{dis}}^l$ on the slab with a positive and a negative values of $d$, respectively, where the sublattice index $l$ is chosen in order that the axis passing through the center of the tetrahedron and the sublattice $l$ become perpendicular to the slab.
We plot the band energies along the lines that connect the four two-dimensional TRIM as shown in Fig. \ref{fig:slab_all}(b).
The spectrum of the gapless surface states are clearly seen for both of $d=0.1$ and $d=-0.1$, at $E/t \sim 0$, namely around the Fermi level at half filling whereas at $E/t \sim -2$ only for positive $d$.
This confirms the above arguments.

\section{Interaction Effects by Hartree-Fock Mean-Field Calculation}\label{sec:Interaction Effects by Hartree-Fock Mean-Field Calculation}

In this section, we study electron correlation effects without the SO interaction.
We show results of Hartree-Fock mean-field calculations of the extended Hubbard model with the obtained phase diagram under competitions among various phases. 
Our result shows that, even without an explicit SO interaction, the form identical to eq. (\ref{eq:HSO}) emerges as a consequence of a spontaneous symmetry breaking in an appropriate range of the interaction parameters.
It leads to the emergence of the STI.
We first concentrate on the half-filled case here.

\subsection{Extended Hubbard model and Hartree-Fock approximation}

The extended Hubbard model under consideration takes the form
\begin{equation}
	H = H_{0} + U\sum_{i}n_{i\uparrow}n_{i\downarrow} + V\sum_{\langle ij \rangle }n_{i}n_{j},
\end{equation}
where $n_{i\sigma} = c_{i\sigma}^{\dagger}c_{i\sigma}$, $n_{i} = n_{i\uparrow}+n_{i\downarrow}$ with $n_{i\sigma}$ being the density operator $n_{i\sigma} \equiv c_{i\sigma}^{\dagger}c_{i\sigma}$ and $\langle ij \rangle$ is the summation over the nearest neighbor pairs.
The onsite interaction is proportional to $U$ and the nearest neighbor interaction is given by $V$.
We decouple the on-site interaction according to
\begin{eqnarray}
	n_{i\uparrow}n_{i\downarrow} &\approx& n_{i\uparrow}\langle n_{i\downarrow}\rangle + \langle n_{i\uparrow}\rangle n_{i\downarrow} - \langle n_{i\uparrow}\rangle \langle n_{i\downarrow}\rangle \notag \\
	&-& c_{i\uparrow}^{\dagger} c_{i\downarrow}\langle c_{i\downarrow}^{\dagger} c_{i\uparrow}\rangle - \langle c_{i\uparrow}^{\dagger} c_{i\downarrow}\rangle c_{i\downarrow}^{\dagger} c_{i\uparrow}  \notag \\
	&+& \langle c_{i\uparrow}^{\dagger} c_{i\downarrow}\rangle \langle c_{i\downarrow}^{\dagger} c_{i\uparrow} \rangle , \label{eq:approxU}
\end{eqnarray}
and the next neighbor interaction as
\begin{eqnarray}
	n_{i}n_{j} &\approx& n_{i}\langle n_{j}\rangle + \langle n_{i}\rangle n_{j} - \langle n_{i}\rangle \langle n_{j}\rangle \notag \\
	&-& \sum_{\alpha\beta} \left( c_{i\alpha}^{\dagger}c_{j\beta}\langle c_{j\beta}^{\dagger}c_{i\alpha}\rangle + \langle c_{i\alpha}^{\dagger}c_{j\beta}\rangle c_{j\beta}^{\dagger}c_{i\alpha} \right. \notag \\
	&-& \left. \langle c_{i\alpha}^{\dagger}c_{j\beta}\rangle \langle c_{j\beta}^{\dagger}c_{i\alpha}\rangle \right). \label{eq:approxV1}
\end{eqnarray}
We note that when the system is a semimetal, where no symmetry is broken, $\langle n_{i\uparrow}\rangle = \langle n_{i\downarrow}\rangle , \langle c_{i\uparrow}^{\dagger}c_{j\uparrow}\rangle = \langle c_{i\downarrow}^{\dagger}c_{j\downarrow}\rangle \neq 0$, and $\langle c_{i\uparrow}^{\dagger}c_{j\downarrow}\rangle = \langle c_{i\downarrow}^{\dagger}c_{j\uparrow}\rangle = 0$ are satisfied.

\subsection{Symmetry of pyrochlore lattice}\label{subsec:Symmetry of pyrochlore lattice}

Before listing up candidate solutions of our mean-field calculations, we summarize the space group of the pyrochlore lattice, which will be useful to identify the mean-field Hamiltonians of the candidate states introduced in the next subsection.
The space group of the pyrochlore lattice is $Fd\overline{3}m$, which contains 24 symmorphic elements of tetrahedral point group $\overline{4}3m$ and 24 non-symmorphic elements.
The non-symmorphic elements, which include inversion, can all be described by a combination of the translation, tetrahedral point group, and inversion.
Therefore, we only list the 24 elements of the tetrahedral point group $T_{\mathrm{d}}$, by taking the position of one tetrahedron's center fixed.
We choose this tetrahedron consisting of the sites 1,2,3,4 in Fig. \ref{fig:pyrochlore}(a).
Its elements are:
\begin{enumerate}
\item 1: The identity
\item $C_{i}^{n} (i=1,2,3,4\ n=1,2)$: $2n\pi/3$ rotation around the axis passing through the site $i$ and the center of the tetrahedron.
\item $C_{a} (a=x,y,z)$: $\pi$ rotation around the axis which passes through the center of the tetrahedron and is parallel to the $a$ axis.
\item $\sigma_{\mathrm{d}ij} (i\neq j\ i,j=1,2,3,4)$: Mirror reflection about the plane determined by the sites $i$, $j$ and the center of the tetrahedron.
\item $s_{a}^{n} (a=x,y,z\ n=\pm1)$: Combination of $n\pi/2$ rotation around the axis parallel to the $a$ axis which passes through the center of the tetrahedron and mirror reflection about the plane $a=1/2$ 
\end{enumerate}
Then the invariant group of the $H_0$ is described by
\begin{equation}
	G_{0} = T\times T_{\mathrm{d}} \times \{1,P\} \times SU(2) \times \{1,\Theta\},
\end{equation}
where $T$ is the translation of the fcc lattice, $P$ is the inversion and $\Theta$ is the time reversal.
The invariant group of a candidate phase must be a subgroup of $G_{0}$.

Now we consider the invariant group of the Hamiltonian when the SO interaction is taken in to account.
Since the interaction is coupling of the orbital and spin angular momentum, it preserves the inversion and time-reversal symmetry.
On the other hand, the elements of $T_{\mathrm{d}}$ must combine with a specific element of $SU(2)$, since the elements of the invariant group must rotate orbital and spin angular momentum equivalently.
To further discuss the symmetry, we define the spin rotation operator $u(\bm{n},\theta)$ which rotates the spins by $\theta$ around $\bm{n}$ axis.
Then the invariant group $G_{\mathrm{SO}}$ is described by
\begin{equation}
	G_{\mathrm{SO}}=T \times \{1,P\} \times L_{\mathrm{SO}}, \label{eq:GSO}
\end{equation}
where $L_{\mathrm{SO}}$ contains
\begin{eqnarray}
	L_{\mathrm{SO}} &=& \{1,\Theta\} \nonumber \\
	&\times& \begin{Bmatrix}
		1 & \\
		C_{i}^{n}u(\bm{n}_{i},\frac{2}{3}n\pi) & (i=1,2,3,4 \ n=1,2) \\
		C_{a}u(\bm{e}_{a},\pi) & (a=x,y,z) \\
		\sigma_{\mathrm{d}ij}u(\bm{n}_{kl},\pi) & (\{i,j,k,l\}=\{1,2,3,4\}) \\
		s_{a}^{n}u(\bm{e}_{a},-\frac{n}{2}\pi) & (a=x,y,z \ n=\pm1)
	\end{Bmatrix}. \nonumber \\
\end{eqnarray}
Here, $\bm{n}_{i}$ is defined by
\begin{eqnarray}
	\bm{n}_{1} &=& \frac{1}{\sqrt{3}}(1,-1,1),\ \bm{n}_{2}=\frac{1}{\sqrt{3}}(-1,1,1), \nonumber \\
	\bm{n}_{3} &=& \frac{1}{\sqrt{3}}(1,1,-1),\ \bm{n}_{4}=\frac{1}{\sqrt{3}}(-1,-1,-1) \label{eq:allout},
\end{eqnarray} 
$\bm{n}_{ij}$ is parallel to the bond connecting the sites $i$ and $j$ on the same unit cell and $\bm{e}_{a}$ is a unit vector of $a$ direction.
Under this symmetry, the form of the Hamiltonian is restricted to $H = H_{0} + H_{SO}$ defined by eqs.(\ref{eq:NNhopping}) and (\ref{eq:HSO}) as long as we consider the nearest neighbor hopping.

\subsection{Candidate phases}\label{subsec:Candidate phases}

Let us now introduce the phases which we find stable at half filling. 
Besides the topologically nontrivial phase, we also examine possible spin density waves (SDW) and charge density waves (CDW).
In our mean-field calculations, we restrict to the phases which do not break the translational symmetry of the fcc lattices.

\subsubsection{Semimetal}
For relatively weak interactions, the semimetal phase arising from $H_0$ remains stable. This phase retains all the symmetries of the Hamiltonian $H_0$.

\subsubsection{Charge-density wave}
\begin{figure}
	\centering
	\includegraphics{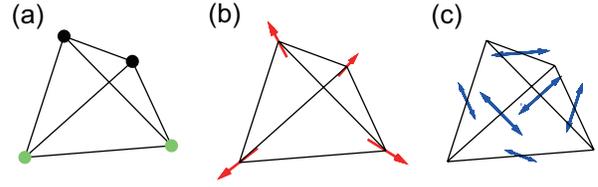}
	\caption{(color online). (a) Unit cell of CDW phase. Black and light (green) sites stand for charge-rich and charge-poor sites, respectively. (b) Unit cell of SDW phase. The arrows denote the magnetization on each site. (c) Unit cell with arrows indicating directions of $\bm{b}_{ij}\times\bm{d}_{ij}$. These are relevant to the Fock terms of the SDW phase and the TMI phase.}
	\label{fig:candidate}
\end{figure}

A CDW phase, where the expectation values of the particle number differ by sites, is expected in the limit of strong nearest neighbor repulsion $V$.
Assuming that the full translational symmetry of the fcc lattice is intact, we find that the CDW phase with two charge-rich sites and two charge-poor sites per unit cell is energetically favored.
When we define $\rho_i$ with $\langle n_{i\sigma}\rangle =f+\rho_{i}$, with $f$ being the filling factor, the phase is, for instance, characterized by
\begin{equation}
	\rho_{1}=\rho_{2}=\rho,\ \rho_{3}=\rho_{4}=-\rho,
\end{equation}
where $\rho$ is the order parameter of the CDW phase.
The unit cell is schematically shown in Fig. \ref{fig:candidate}(a).
The symmetry of this state is described by
\begin{equation}
	G_{\mathrm{CDW}}=T \times \{1,C_{z},\sigma_{\mathrm{d}12},\sigma_{\mathrm{d}34}\} \times \{1,P\} \times SU(2) \times \{1,\Theta\}.
\end{equation}
This symmetry allows the bond expectation values to have three degrees of freedom, which are $\langle c_{1\sigma}^{\dagger}c_{2\sigma}\rangle ,\ \langle c_{3\sigma}^{\dagger}c_{4\sigma}\rangle $, and remaining 4 equivalent bonds.
When we define $g_{ij} = \langle c_{j\sigma}^{\dagger}c_{i\sigma}\rangle$, the mean-field Hamiltonian takes the form
\begin{eqnarray}
	H_{\mathrm{MF\_CDW}} &=& H_{\mathrm{0}} + \sum_{i}(U-4V)\rho_{i}n_{i} \nonumber \\
	&-& V\sum_{\langle ij \rangle \sigma}(g_{ij} c_{i\sigma}^{\dagger}c_{j\sigma} + h.c.) \nonumber \\
	&+& \left(-4U\rho^{2}+4V\rho^{2} \right. \nonumber \\
	&+& \left. 4Vg_{12}^{2}+4Vg_{34}^{2}+16Vg_{13}^{2}\right)L^{3}, \label{eq:HMF_CDW}
\end{eqnarray}
where $L^{3}$ is the number of the unit cells.
We note that $g_{13}=g_{14}=g_{23}=g_{24}$ and $g_{ij} = g_{ji}$ is satisfied under this symmetry.

\subsubsection{Topological Mott insulator}
TMI phase is favored at an intermediate strength of $U$ and $V$, similarly to the QSH phases in two dimensions \cite{cit:PRL100_156401,cit:PRB82_075125} and the TMI phase in the diamond lattice \cite{cit:PRB79_245331}.
As mentioned above, the invariant subgroup of $H_{0} + H_{\mathrm{SO}}$ is given by eq. (\ref{eq:GSO}).
The bond expectation value corresponding to this symmetry is described by
\begin{equation}
	\langle c_{j\beta}^{\dagger}c_{i\alpha}\rangle = \left( g\sigma_{0} - i\sqrt{2}z\frac{\bm{b}_{ij}\times\bm{d}_{ij}}{|\bm{b}_{ij}\times\bm{d}_{ij}|}\cdot\bm{\sigma} \right)_{\alpha\beta}, \label{eq:TMI_bond}
\end{equation}
where nonzero $z$ reduces the invariance subgroup of the system from $G_{0}$ to $G_{\mathrm{SO}}$.
The order parameter $z$ is explicitly written as
\begin{equation}
	z = \frac{i}{2\sqrt{2}}\sum_{\alpha\beta}\langle c_{j\beta}^{\dagger}c_{i\alpha}\rangle \left(\frac{\bm{b}_{ij}\times\bm{d}_{ij}}{|\bm{b}_{ij}\times\bm{d}_{ij}|}\cdot\bm{\sigma} \right)_{\beta\alpha}.
\end{equation}
On the other hand, $g$ does not break any symmetry of the system.
The mean-field Hamiltonian for the present case is
\begin{eqnarray}
	H_{\mathrm{MF\_TMI}} &=& H_{\mathrm{0}} 
	- Vg\sum_{\langle ij\rangle \sigma}(c^{\dagger}_{i\sigma}c_{j\sigma}+h.c.) \nonumber \\
	&+& \sqrt{2}Vz\sum_{\langle ij\rangle \alpha\beta}\left( i c_{i\alpha}^{\dagger}\frac{\bm{b}_{ij}\times\bm{d}_{ij}}{|\bm{b}_{ij}\times\bm{d}_{ij}|}\cdot\bm{\sigma}_{\alpha\beta}c_{j\beta} + h.c. \right) \nonumber \\
	&+& (24Vg^2 + 48Vz^2)L^3 \label{eq:HMF_TMI}.
\end{eqnarray}
We note that the energy remains invariant under the global spin rotation and directly leads to an $SU(2)$ degeneracy.
When $z \neq 0$, the invariance subgroup of $H_{\mathrm{MF\_TMI}}$ is described by $G_{\mathrm{TMI}} = G_{\mathrm{SO}}$.
As mentioned above, the system becomes the STI for the Hamiltonian eq. (\ref{eq:NNhopping}) and (\ref{eq:HSO}).
Therefore the ground state of the Hamiltonian also is the STI when the solution with $z \neq 0$ is stabilized.

\subsubsection{Spin density wave}
When $U$ becomes larger, antiferromagnetic phases may be stabilized.
The phase is characterized by the order parameter $\Delta$ with, for instance, $\langle \bm{S}_{i}\rangle = \Delta\bm{n}_{i}$, where $\bm{S}_{i} = c_{i\alpha}^{\dagger}\bm{\sigma}_{\alpha\beta}c_{i\beta}/2$ is the local spin operator and $\bm{n}_{i}$ is defined by eq. (\ref{eq:allout}) for $i=1,2,3,4$.
This is nothing but the ``all-out" phase illustrated in Fig. \ref{fig:candidate}(b).
This phase breaks the time-reversal symmetry and the spin rotational symmetry.
The invariant subgroup of this phase is described by
\begin{equation}
	G_{\mathrm{SDW}}=T \times \{1,P\} \times L_{\mathrm{SDW}}.
\end{equation}
Here $L_{\mathrm{SDW}}$ contains
\begin{equation}
	L_{\mathrm{SDW}} = \begin{Bmatrix}
		1 & \\
		C_{i}^{n}u(\bm{n}_{i},\frac{2}{3}n\pi) & (i=1,2,3,4 \ n=1,2) \\
		C_{a}u(\bm{e}_{a},\pi) & (a=x,y,z) \\
		\sigma_{\mathrm{d}ij}u(\bm{n}_{kl},\pi)\Theta & (\{i,j,k,l\}=\{1,2,3,4\}) \\
		s_{a}^{n}u(\bm{e}_{a},-\frac{n}{2}\pi)\Theta & (a=x,y,z \ n=\pm1)
	\end{Bmatrix}.
\end{equation}
Using this symmetry, the form of the Fock terms is restricted to
\begin{eqnarray}
	\langle c_{j\beta}^{\dagger}c_{i\alpha}\rangle &=& \left( g\sigma_{0} + \eta\frac{\bm{n}_{i}+\bm{n}_{j}}{|\bm{n}_{i}+\bm{n}_{j}|}\cdot\bm{\sigma} \right. \notag \\
	&-& \left. i\sqrt{2}z\frac{\bm{b}_{ij}\times\bm{d}_{ij}}{|\bm{b}_{ij}\times\bm{d}_{ij}|}\cdot\bm{\sigma} \right)_{\alpha\beta},
	\label{eq:Fock_SDW}
\end{eqnarray}
where $g$, $\eta$ and $z$ are order parameters.
The order parameter $z$  is nothing but that appeared in eq. (\ref{eq:TMI_bond}) and this term must be included since nonzero $z$ preserves the invariant subgroup $G_{\mathrm{SDW}}$.
This indicates $G_{\mathrm{SDW}} \subset G_{\mathrm{TMI}}$ is satisfied.
We note that nonzero $\Delta$ or $\eta$ reduces the invariance group of the system to $G_{\mathrm{SDW}}$.
The mean-field Hamiltonian of the SDW takes the form
\begin{eqnarray}
	H_{\mathrm{MF\_SDW}} &=& H_{\mathrm{0}} -U\Delta\sum_{i\alpha\beta}c_{i\alpha}^{\dagger}(\bm{n}_{i}\cdot\bm{\sigma})_{\alpha\beta}c_{i\beta} \notag \\
	&-& Vg\sum_{\langle ij\rangle \sigma}(c^{\dagger}_{i\sigma}c_{j\sigma}+h.c.) \notag \\
	&-& V\sum_{\langle ij\rangle\alpha\beta}\left( \eta c_{i\alpha}^{\dagger}\frac{\bm{n}_{i}+\bm{n}_{j}}{|\bm{n}_{i}+\bm{n}_{j}|}\cdot\bm{\sigma}_{\alpha\beta}c_{j\beta} + h.c. \right) \notag \\
	&+& V\sum_{\langle ij\rangle\alpha\beta}\left( i\sqrt{2}z c_{i\alpha}^{\dagger}\frac{\bm{b}_{ij}\times\bm{d}_{ij}}{|\bm{b}_{ij}\times\bm{d}_{ij}|}\cdot\bm{\sigma}_{\alpha\beta}c_{j\beta} + h.c. \right) \notag \\
	&+& (4U\Delta^2 + 24Vg^2 + 24V\eta^2 + 48Vz^2)L^3.
\end{eqnarray}

\subsection{Phase diagram}\label{subsec:The phase diagram}
\begin{figure}
	\centering
	\includegraphics{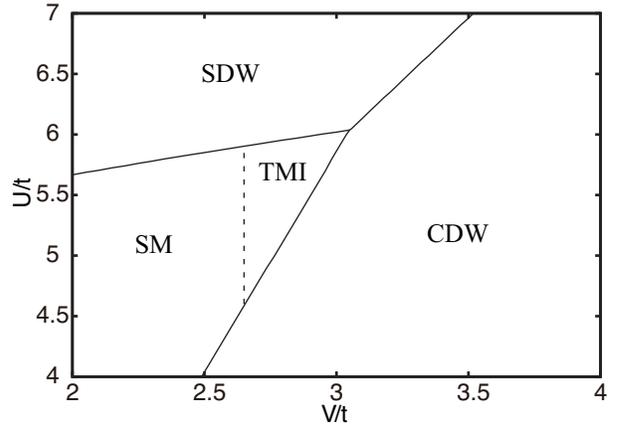}
	\caption{Phase diagram of extended Hubbard model on pyrochlore lattice. The solid (dashed) lines indicate first order (continuous) transitions. SM, TMI, CDW, and SDW denote semimetal, topological Mott insulator, charge density wave and spin density wave, respectively.}
	\label{fig:phase_diagram}
\end{figure}

We have searched the global minimum with respect to its order parameters and compared energies among different possible phases.
The $U-V$ phase diagram is shown in Fig. \ref{fig:phase_diagram}.
The semimetal phase exists at a small $U$ and $V$ region.
When $U$ is dominant, the SDW is stable and the transition from the semimetal to the SDW is of the first order.
The CDW is stabilized when $V$ is dominant and the transition from the semimetal to the CDW is also of the first order.
Most importantly, we find the TMI phase at intermediate $U$ and $V$, where the transition from the semimetal is continuous.
The transition between the semimetal and TMI occurs at $V/t \sim 2.6$, and the order parameter $z$ grows continuously from that point.
Other details such as critical exponents are beyond the scope of this paper, and would be future issues.

The phases at other fillings are also the subject of interest.
At quarter filling, we find CDW phases with three equivalent charge-poor sites per unit cell for relatively large $V$ and a ferromagnet for large $U$ limit.
The system is a semimetal for relatively small $U$ and $V$.
We note that the topologically nontrivial states are not seen even as a local minimum.
At three-quarter filling, the state without a symmetry breaking is unstable due to the divergence of the density of states at the Fermi level.
We find ferromagnet and CDW with three equivalent charge-rich sites per unit cell when $U$ and $V$ dominates, respectively, whereas we do not find the topological insulator phase.

\subsection{Discussion}

We first discuss the possibility of experimental realization of the TMI along with the present mechanism.
The calculations in $\S$\ref{subsec:Candidate phases} and \ref{subsec:The phase diagram} are for $T=0$, while we can treat the system at nonzero temperatures as well in the grand canonical ensemble.
For $V/t \sim 3.0$ we find that the transition to TMI occurs at $T_{c}\sim 0.01t$.

To realize the TMI state in experiments, we propose that compounds like $\mathrm{Cd_{2}Os_{2}O_{7}}$ and $\mathrm{Tl_{2}Ru_{2}O_{7}}$ are the candidates since their band structures near the Fermi level resemble that of our present model \cite{cit:PRB63_195104,cit:JPSJ69_526}.
Especially in $\mathrm{Tl_{2}Ru_{2}O_{7}}$, it is reported in Ref.\citen{cit:JPSJ69_526} that bands which consist of antibonding states of Tl $6s$ and O $2p$ orbitals exist near half filling.
The paper also shows a calculated band structure of the tight-binding model for Tl $6s$ orbitals.
Though the hopping parameter of the model is not given in the paper, we estimate $t\sim 0.3$ eV from the band width.
Then the critical temperature to TMI is $T_{c}\sim 3 \times 10$ K, which is accessible in experiments.
To realize the TMI state in such materials, we have to change the Fermi level and control the parameters in an extended Hubbard model.
We propose that an electron doping as well as application of pressure are effective for this purpose.

Compared to other lattice models which result in the topological insulators at the Hartree-Fock mean-field level, we propose that the pyrochlore lattice is the best candidate for the experimental realization for the following two reasons.
First, since the tight-binding model on the pyrochlore lattice becomes the topological insulators solely with the nearest neighbor hopping, we do not need a strong next nearest neighbor Coulomb repulsion that had been needed in other cases \cite{cit:PRL100_156401,cit:PRB79_245331}.
Second, the transition is possible at $T \neq 0$, because our model is three dimensional in space in contrast to the kagom\'{e} lattice.

It is interesting to further ask what kind of interaction enhances the stability of the TMI state in the competitions with the other phases.
Here, we show that the ferromagnetic interaction of the form
\begin{equation}
	H_{\mathrm{FM}}=-J\sum_{\langle ij\rangle}\bm{S}_{i}\cdot\bm{S}_{j},
\end{equation}
suits for this purpose.
This can be easily confirmed by calculating coefficients of the order parameters of the mean-field Hamiltonians.
The coefficients of the order parameters in eq. (\ref{eq:HMF_TMI}) change as
\begin{equation}
	Vg \rightarrow \left(V-\frac{3}{4}J\right)g, \ Vz \rightarrow \left(V+\frac{1}{4}J\right)z,
\end{equation}
from which we see that the interaction stabilizes the TMI phase since it is characterized by a nonzero $z$ when the time-reversal symmetry is intact.
On the other hand, the ferromagnetic interaction destroys the SDW state, because it is antiferromagnetic, and does not make difference from the CDW state, because the coefficient of $\rho$ in eq. (\ref{eq:HMF_CDW}) does not change.

In the parameter space of $V$ and $\lambda$, the Hamiltonian between SM and TMI is either continuous or first order.
In the TMI phase, the spontaneous symmetry breaking takes place at $\lambda = 0$.
On the other hand, the topology change is also involved at the transition.
The interplay of topology change and the symmetry breaking is an intriguing future issue in terms of theory of quantum criticality \cite{cit:PRB72_075113,cit:PRB75_115121}.

\section{Conclusion and Summary}

In the first part of this paper, we have presented microscopic tight-binding models on the pyrochlore lattice which exhibit topological insulator phases.
It has turned out that, in tight binding models, the nearest neighbor transfer is sufficient to realize the topological insulator phase without relying on transfers between pairs at further distance, which is, as far as we know, the first case in 3D systems.
We have also shown that the nearest neighbor transfer to realize the topological insulator phase exists when SO interaction is intact.
The $Z_2$ index is $(1;000)$ at half filling, whereas at quarter filling, the system is on the critical point separating STI and WTI when distortions are absent.
When we view the pyrochlore lattice as an alternative stacking of kagom\'{e} and triangle lattice, distortions may either enhance or weaken interlayer hopping transfer.
According as it, the distortion may make the system either STI or WTI.
We have also confirmed the topological nontriviallity of the system by showing the presence of the surface states on the slab geometry, which is consistent with the $Z_2$ index.

In the latter part of this paper, we have studied the extended Hubbard model on the pyrochlore lattice.
We have examined candidate states at half filling from group theoretical arguments in the mean-field Hamiltonians.
The numerical Hartree-Fock mean-field calculations show that the topological insulator phase is energetically favored for an appropriate range of parameters with the nearest neighbor Coulomb repulsion.
The transition from the semimetal to the TMI phase is continuous, while transitions to CDW and SDW are of the first order.

In addition, we have also found that the ferromagnetic interaction enhances the stability of the topological state.
The calculation in the grand canonical ensemble has shown that the critical temperature to the topological phase is as much as 0.01t and is experimentally accessible.

{\bf Acknowledgements}
The authors thank Yuto Ito and Takahiro Misawa for fruitful discussions. This work is financially supported from MEXT Japan under the grant numbers 22104010 and 22340090.

\end{document}